# Magnetization Measurement of a Possible High-Temperature Superconducting State in Amorphous Carbon Doped with Sulfur

Israel Felner[1] and Yakov Kopelevich[2]

[1]Racah Institute of Physics, The Hebrew University, Jerusalem, 91904, Israel

[2]Instituto de Física "Gleb Wataghin", Universidade Estadual de Campinas - UNICAMP

13083-970, Campinas, SP, Brasil

Abstract

Magnetization M(T,H) measurements performed on thoroughly characterized commercial amorphous carbon powder doped with sulfur (AC-S), revealed the occurrence of an inhomogeneous superconductivity (SC) below $T_c$ = 38 K. The constructed magnetic field-temperature (H-T) phase diagram resembles that of type-II superconductors. However, AC-S demonstrates a number of anomalies. In particular, we observed (1) a non-monotonic behavior of the lower critical field $H_{c1}(T)$; (2) a pronounced positive curvature of the "upper critical field boundary" that we associated with the flux lattice melting line $H_m(T)$; (3) a spontaneous ferromagnetic-like magnetization $M_0$ coexisting with SC. Based on the analysis of experimental results we propose a nonstandard SC state in AC-S.

Recently, possible superconductivity (SC) in graphene (isolated graphitic layer) has attracted a considerable theoretical attention [1-9]. In many respects, this was motivated by the enormous current interest of graphene itself [10-12], and by observations of SC at elevated temperatures in related materials: doped fullerenes [13, 14] graphite [15-19], diamond [20, 21], and carbon nanotubes [22, 23]. In particular, BCS type SC in graphene with a mean-field critical temperature $T_c^{MF}$ up to ~ 150 K was calculated in Refs. [5, 7], and $T_c^{MF}$ well above the room temperature was predicted within a framework of the resonating valence bond (RVB) model [1, 8], originally proposed by Anderson [24] for high-$T_c$ cuprates. From the experimental side, localized SC found for graphite-sulfur (G-S) composites [15-17] is perhaps the most suggestive realization of the theoretical expectations for the doping-induced SC in graphene [1, 2, 4, 8]. Both d-wave [1, 8] and p-wave [1, 25, 26] symmetries of the SC order parameter (OP) were predicted for graphene. The occurrence of p-wave SC in G-S is appealing, because it coexists with the ferromagnetism (FM) [27], and the interaction between SC and FM OPs has been experimentally demonstrated [17]. According to Ref. [25], high-$T_c$ p-wave SC emerges in a matrix of curved graphene layers with inserted pentagons and heptagons. If high-$T_c$ p-wave SC exists in graphitic materials, this may have far reaching consequences, due to the non-Abelian statistics of vortices in p-wave SC, allowing for the quantum computation [28]. Hence, the research in this direction has a broad and an interdisciplinary interest. Amorphous carbon (AC) is a strongly disordered material consisting of a submicron curved graphene layers with a mixed interlayer stacking. AC also contains partially graphitized carbon fragments that possess both negative and positive curvatures, required for SC [29]. All these motivations triggered the present work.

This report is focused on experimental evidences for the SC state in sulfur-doped amorphous carbon (AC-S) occurring at T < $T_c$ = 38 K. The observed anomalous behavior of lower and apparent upper critical fields, as well as the coexistence of superconducting and ferromagnetic-like states, all suggest an unconventional superconducting state in AC-S.

The pristine material was a 75 years old commercial amorphous carbon powder manufactured by Fisher (C190-N) as decolorized carbon. AC and sulfur (99.998%; Aldrich Chemical Company, Inc.) powders were mixed in a mass ratio $m_C:m_S$ = 2:1, pressed into pellets, sealed in evacuated quartz tube and then heated at 250 °C for 24 hours before cooling down to ambient temperature.

The samples were thoroughly characterized by means of x-ray diffraction (XRD), scanning electron microscopy (SEM), energy dispersive spectroscopy (EDS) JOEL JSM-7700 SEM and $^{57}$Fe Mossbauer spectroscopy. Trace element analysis was performed by means of the inductively coupled plasma mass spectrometer (ICPMS) (Perkin-Elmer ICP-OES model 3300) of acid extracts. M(T, H) measurements up to H = 50 kOe and 5 K ≤ T ≤ 300 K were performed by using commercial (MPMS5 Quantum Design) SQUID magnetometer. The zero-field-cooled (ZFC) magnetization $M_{ZFC}$(T,H) was measured on heating the sample after cooling at H = 0. Prior to each such measurement, the SQUID was adjusted to compensate the remnant magnetic field of the SC solenoid. The field-cooled data $M_{FCC}$(T,H) were taken under cooling in applied field.

The featureless XRD patterns obtained for both AC and AC-S samples are consistent with their amorphous-like structure. From SEM images, we found a broad distribution in the carbon "grain" size which ranges from ~ 10 nm to several microns. Spatially resolved elemental composition analysis performed by EDS yields: Na (0.30(1) at%), oxygen (2.44(1) at%) and sulfur (0.21(1) at%) as extra elements. The analysis performed on AC-S sample showed: Na (0.35 at%), oxygen (1.96 at%) and sulfur (10.3 at%). Thus, the major difference between the AC and AC-S samples is the significant increase of the sulfur contents in AC-S. The mass ratio $m_C$:$m_S$ are 174 and 3 for AC and AC-S samples, respectively. The trace element analysis of AC revealed impurities (ppm): V (2.05), Ni (2.77), Zn (7.09), Cu (11.1), Mn(133.1), Al (212.7), Fe (360.0), and Na (4625). It appears that the total amount of magnetic impurities (Ni, Mn and Fe) is about 486 ppm. Long-time (two weeks) room temperature $^{57}$Fe Mossbauer measurements, performed on AC sample, revealed a broad magnetic spectrum with two sextets, with an estimated Fe concentration of ~ 350 ± 50 ppm. A least square fit provides evidence that the magnetic sextets are related to magnetite ($Fe_3O_4$). From M(H) measurements of AC, we deduced a spontaneous FM magnetization of $M_m \approx$ 0.033 emu/g that coincides with the value corresponding to 350 ppm of $Fe_3O_4$ ($M_S$ = 94.5 emu/g) obtained from Mossbauer measurements.

Figure 1 (a) shows $M_{ZFC}$(T, H) and $M_{FCC}$(T, H) curves measured for pristine AC sample at H = 22 Oe. Fig. 1(b) presents $M_{ZFC}$(T, H) and $M_{FCC}$(T, H) obtained for H = 50 Oe, as well as the remnant magnetization $M_{REM}$ (T, H = 0), recorded after the FCC process when the field was switched off at T = 5 K (here, the irrelevant temperature-independent background magnetization $M_m$ = 0.037 emu/g measured for this sample was subtracted). As Fig. 1(b) demonstrates, at T < $T_c$ ~ 38 K, the magnetization is strongly irreversible and $M_{FCC}$(T) > $M_{ZFC}$(T). Both $M_{ZFC}$(T) and $M_{REM}$(T) show a pronounced step-like feature at $T_c$ ~ 38 K. Importantly, $M_{FCC}$(T) also demonstrates a clear drop (though, smaller in amplitude) below $T_c$. It appears, that $M_{ZFC}$(T) is negative (diamagnetic) below $T_c$, as expected for SC, where the diamagnetism originates from screening supercurrents, and the drop of $M_{FCC}$(T) is associated with the magnetic flux expulsion due to Meissner effect (ME). Then, it is reasonable to relate $M_{REM}(T)$ to the trapped magnetic flux (vortices). We estimate the shielding fraction deduced from $M_{ZFC}(T)$ to be ~ 0.15 %, and the much smaller Meissner fraction (MF) ~ 0.02%. Since no corrections for carbon magnetism (see below) and for the flux trapping effects were made, the ME provides only the lower limit for the SC volume fraction. Nevertheless, the smallness of both shielding and MF values, suggests an inhomogeneous character of SC in AC-S. The data presented in Fig. 1 (a), indicate also the possible SC with $T_c$ = 32 K in pristine sample, giving the SC shielding fraction of ~ 0.03 %, i. e. 5 times smaller than that obtained for AC-S. The much smaller SC fraction in AC would explain the invisibility of the Meissner signal, as $M_{FCC}(T)$ in Fig. 1(a) shows. The comparison of the results obtained for AC and AC-S samples indicates that the sulfur enhances both $T_c$ and the SC fraction.

Figure 2 (a, b) shows the normalized ZFC magnetization $M_{ZFC}(T)/M_{ZFC}(40\ K)$ measured for various fields. It can be readily seen that for H = 25 kOe, the SC-like transition is suppressed and (reversible) magnetization demonstrates a paramagnetic (PM) behavior. As Fig. 2(b) exemplifies, $T_c(H)$ at lower fields can be well defined. The phase boundary $T_c(H) \equiv H_m(T)$ which presumably separates SC and normal states is shown in Fig. 3(a). The available experimental points for analysis are limited by t = $T/T_c \geq 0.8$ where the data can be best described by the power law:

$$H_m = H_{m0}(1-T/T_c)^{\alpha}, \qquad (1)$$

where $\alpha$ = 2.4 ± 0.2, $H_{m0}$ = 6.5·10$^5$ Oe, and $T_c$ = 38 K, i. e. $H_m(T)$ demonstrates the positive curvature.

For $H < H_{c1}(T)$, $M_{ZFC}(H)$ isotherms linearly decrease with the field (see the main panel and the upper inset in Fig. 4), as expected for SC in the Meissner state; $M_{ZFC}(H) = M_0 - \chi_d(T)H$, where $\chi_d(T)$ is the absolute value of the diamagnetic susceptibility, and $M_0 \geq 0$ is the spontaneous magnetization whose values depend on a thermo-magnetic sample history (see below). Noting, that $M_0$ does not alter $\chi_d(T)$. Unexpectedly, $|\chi_d(5K)|$ is smaller than that measured at $T = 10$ K, 20 K, and 25 K. For $H > H_{c1}$, $M_{ZFC}(H)$ deviates from the linearity for $H \geq H_{c1}(T)$. The $H_{c1}(T)$ is plotted in Fig. 3(b), where the straight line is obtained from the equation $H_{c1} = H_{c10}(1-T/T_c)$ with $H_{c10} = 550$ Oe. Noting an anomalous drop of $H_{c1}$ found at $T = 5$ K. The lower inset in Fig. 4 presents the $M_{ZFC}(H)$ hystersis loop measured at $T = 5$ K up to $H = 50$ kOe. The PM behavior is evident from the reversible non-saturating high-field portion of the M(H), giving the PM susceptibility $\chi_p \approx 10^{-5}$ emu/g·Oe, that fits well to the known $\chi_p$ values for disordered carbon materials, see e.g. [30]. It should be also emphasized that the $M_{REM}(H = 0, T = 5\ K) \approx 0.24$ emu/g is ~ 7 times bigger than $M_m \approx 0.037$ emu/g ($Fe_3O_4$), i. e. the $M_{REM}$ is essentially related to the AC-S matrix.

Figure 5 sheds light on the origin of $M_0$. Starting with $M_{ZFC}(T, 50\ Oe)$ measurements, after cooling from $T = 300$ K to $T = 5$ K at $H = 0$, the field was switched off at $T = 40 - 43$ K, and $M_{REM}$ (40 K) was recorded. Fig 5(a) describes the measuring procedure for $M_{ZFC}$ (T, 500 Oe). At $T = 43$ K the field was switched off and $M_{REM}$ (point 2) was recorded. Then, the sample was cooled down to $T = 5$ K, and $M_0$ was measured (point 3). After that, higher field were applied (Fig 5, b and c), and the same procedure was repeated. As can be seen, $M_{REM}\ (T > T_c) = M_0(T = 5\ K)$. It also appears, that $M_{REM}\ (M_0) \sim H^\nu$ with $\nu = 0.4$. The occurrence of $M_{REM}$, and its increase with the field are characteristic features of FM. Hence, it is reasonable to relate $M_0$ with the FM magnetization associated with the carbon matrix. However, our measurements do not reveal any signature for FM transition in the entire temperature interval 5 K < T < 275 K, see Fig. 5(d). On the contrary, for $T < T_{max} \approx 150$ K, both $M_{ZFC}(T)$ and $M_{FCC}(T)$ (measured as described earlier) decrease with the temperature decreasing. Noting, (i) $M_{REM} \approx 0.24$ emu/g obtained from M(H) at T = 5 K (Fig. 3, lower inset) agrees well with $M_0(H) \sim H^{0.4}$ dependence, indicating that $M_{REM}$ mainly originates from the FM but not from trapped SC vortices; (ii) heating the sample to room temperature totally suppresses $M_{REM}$. In what follows we speculate on the origin of experimental observations.

In conventional SC, near $T_c$, $H_{c2}(T) \sim (1 - T/T_c)$ for bulk, and $H_{c2}(T) \sim (1 - T/T_c)^{1/2}$ for granular SC [31]. To the best of our knowledge, positive curvature in $H_{c2}(T)$ near $T_c$ appears only in the model of the Bose-Einstein condensation of preformed Cooper pairs [32]. However, in this case $H_{c2}(T) \sim (1 - T/T_c)^{3/2}$ [32]. Alternatively, in the presence of strong thermal and/or quantum fluctuations, melting of Abrikosov vortex lattice (AVL) may occur. In this case, $\alpha \leq 2$, in Eq. (1) is predicted [33]. On the other hand, $\alpha = 2.5$ in Eq.(1) is predicted for the melting of Skyrmion flux (SF) lattice in p-wave SC [34], where Skyrmion is a coreless vortex-like structure carrying two or more flux quanta $\Phi_0$ = h/2e. As Fig. 3 (a) illustrates, the experimental data agree well with this prediction.

In addition, close to $H_{c1}(T)$, all $M_{ZFC}(H)$ isotherms (Fig. 4) can be very well described by the equation:

$$M_{ZFC}(H) = M_0 - \chi_d H + c(H/H_{c1} - 1)^2, \qquad (2)$$

expected for type-II SC, in which the underlying physics is dominated by SF- lattice rather than by AVL [35].

In the chiral p-wave SC, such as $Sr_2RuO_4$, with the OP $p_\pm = p_x \pm ip_y$, domains with $p_-(= p_x - p_y)$ and $p_+(= p_x + p_y)$ states can be formed [36, 37]. If the domain walls are pinned at the sample surface, they may act as "weak links" or channels for vortex entry at $H < H_{c1}$. In general, pinning of any elastic manifold is more effective at lower temperatures. This may explain the drop of $H_{c1}$ at T = 5 K, see Fig. 3(b). We stress, that the reentrance in $H_{c1}(T)$ cannot be easily understood within a framework of classical models.

Next, we discuss the origin of FM-like behavior of AC-S matrix in both "normal" and SC states. We believe that the most plausible (and simplest) explanation is the occurrence of a Griffiths phase, consisting of randomly distributed uncorrelated FM clusters. Such a state has been observed in various inhomogeneous systems, e. g. manganites and cuprates [38, 39]. The rather small exponent $\nu = 0.4$ in $M_0 \sim H^\nu$ supports the Griffiths scenario [39]. We speculate that the FM Griffiths phase coexists with the SF-lattice for $T < T_c(H)$ and with SF-liquid for $T > T_c(H)$. If the SF-liquid state, characterized by enhanced diamagnetism due to formation of Cooper pairs (2e) lacking the phase coherence, extends up to $T_{max} \sim 150$ K, then the magnetization drop below $T_{max}$, see Fig. 5(d), is also understood. In fact, this scenario is in a close analogy with the pseudo-gap phase model for cuprates [40, 41].

In conclusion, we observed SC in sulfur-doped amorphous carbon at $T_c = 38$ K. The obtained results can be consistently understood adopting theoretical predictions for p-wave SC. This work triggers other important questions such as: (i) whether SC in AC-S and in G-S composites is of the same origin? Besides of the fundamental question on the nature of (inhomogeneous) FM in graphitic materials [27], also supported by the present work, (ii) what is the interrelation between FM and SC states? Most importantly, a systematic experimental work, aiming to increase the reproducibility as well as the SC volume fraction in doped AC, is needed.

This work was supported by the Klachky Foundation for Superconductivity, FAPESP and CNPq. Y. K. also gratefully acknowledges the Lady Davis Foundation. We thank I. Nowik for Mossbauer measurements, E. B. Sonin for useful discussions, and Yuval Simons for his assistance.

**FIGURE CAPTIONS**

**Fig. 1. (Color online)** (a) $M_{ZFC}(T)$ and $M_{FCC}(T)$ measured for pristine AC at H = 22 Oe, (b) $M_{ZFC}(T)$ and $M_{FCC}(T)$ measured for AC-S sample at H = 50 Oe, as well as $M_{REM}(T, H = 0)$; the background magnetization $M_m = 0.037$ emu/g is subtracted.

**Fig. 2. (Color online)** (a) Normalized $M_{ZFC}(T)$ plots, measured for H = 50 Oe ( ), 100 Oe (o), 500 Oe ($\Delta$), 1 kOe ($\nabla$), 2.5 kOe ($\Diamond$), 5 kOe (+), 10 kOe (x), 25 kOe (*); (b) the same as in (a) for 3 measuring fields; arrows indicate $T_c(H)$ as defined for H = 100 Oe.

**Fig. 3. (Color online)** (a) Double logarithmic plot of $H_m$ vs $1 - T/T_c$, representing the apparent upper critical field boundary. Dashed line corresponds to Eq. (1); (b) Lower critical field $H_{c1}(T)$ obtained as indicated in Fig. 4 (upper inset); the straight line is the linear fit: $H_{c1} = H_{c10}(1-T/T_c)$, $H_{c10} = 550$ Oe, $T_c = 38$ K.

**Fig. 4. (Color online)** Low-field portions of various $M_{ZFC}(H)$ isotherms after subtraction of the spontaneous ZFC magnetization $M_0(H)$ (see Fig. 5 and the text). The upper inset exemplifies the validity of Eq. (2) that fits $M_{ZFC}(H)$ for T = 5 K (solid line) with $M_0 = 0.04$ emu/g, $H_{c1} = 200$ Oe, and $c = 10^{-3}$. The dotted straight line is $M_{ZFC}(H) = -\chi_d B$, $|\chi_d| = 3.4 \cdot 10^{-5}$ emu/g·Oe. The lower inset shows $M_{ZFC}(H)$ hysteresis loop measured at T = 5 K up to B = 50 kOe.

**Fig. 5.** (a, b, c) $M_{ZFC}(T, H)$ illustrating the appearance of $M_0(H)$ for three measuring fields. All the measurements were made in the sequence $1 \to 2 \to 3$ as indicated in (a); (d) $M_{ZFC}(T)$ and $M_{FCC}(T)$ measured at $H = 50$ Oe in the temperature interval $5\ K \leq T \leq 275\ K$; $T_c = 38$ K and $T_{max} = 150$ K are noted by arrows.

Fig. 1

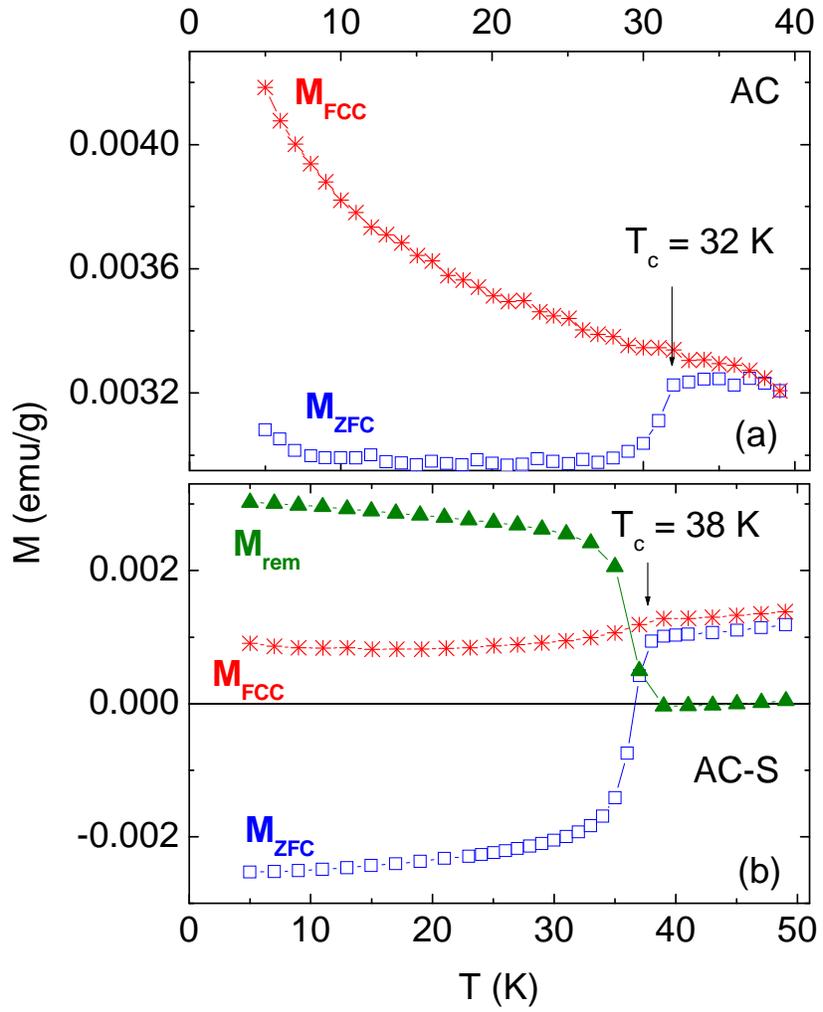

Fig. 2

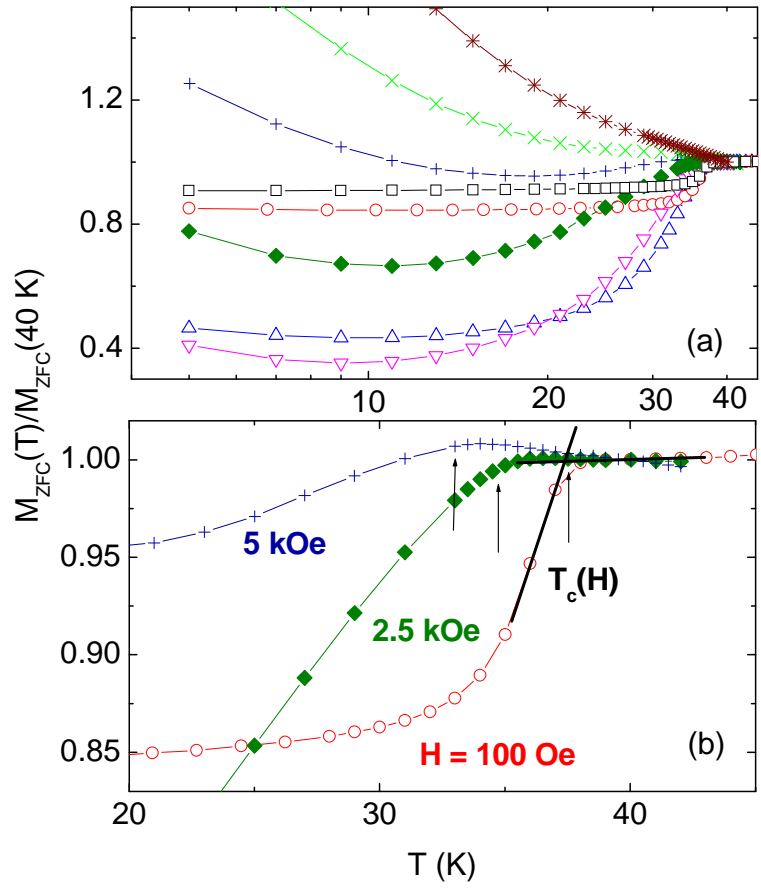

Fig. 3

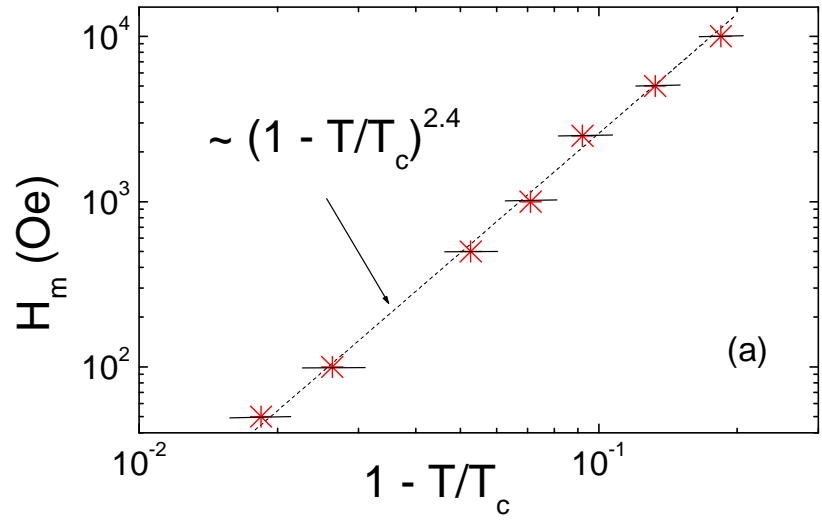

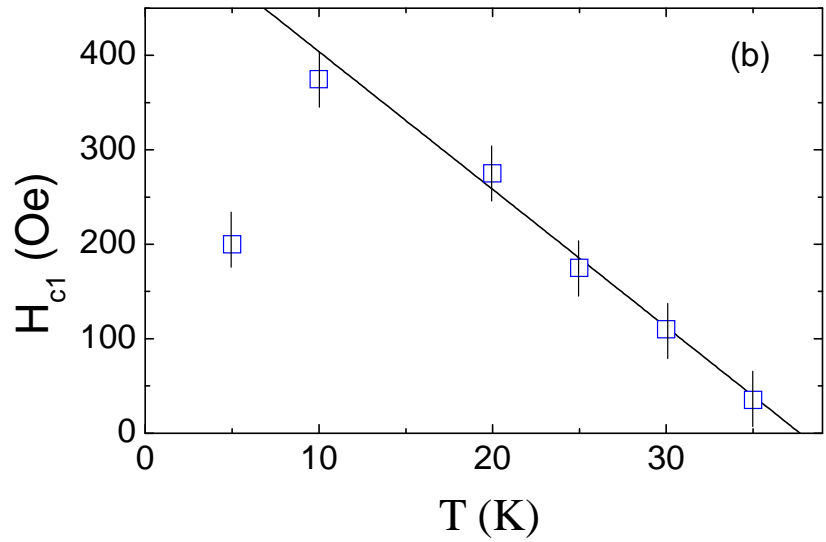

Fig. 4

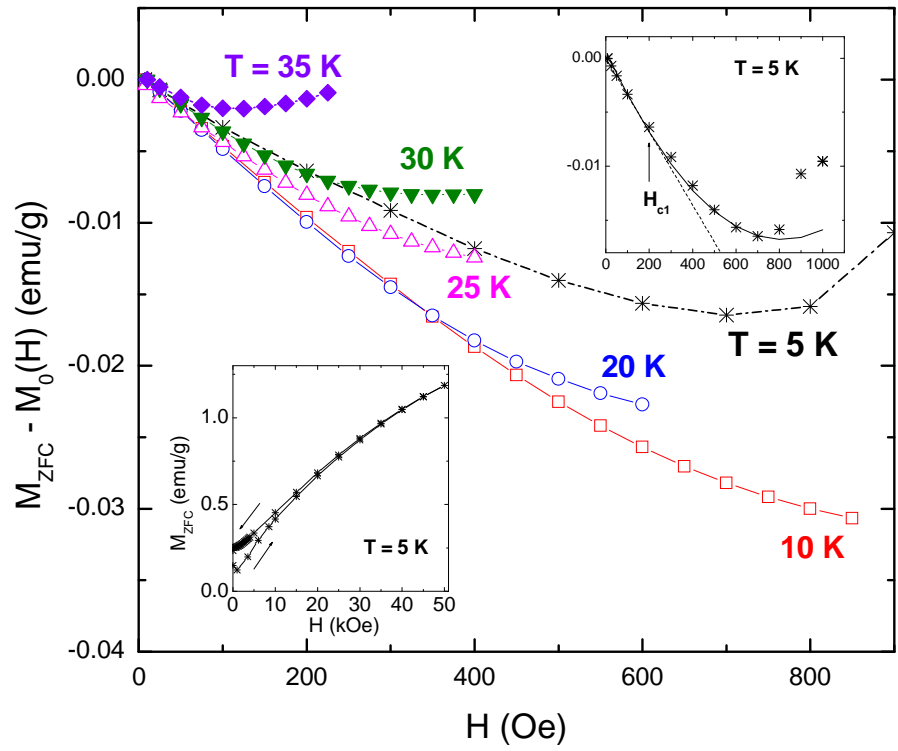

Fig. 5

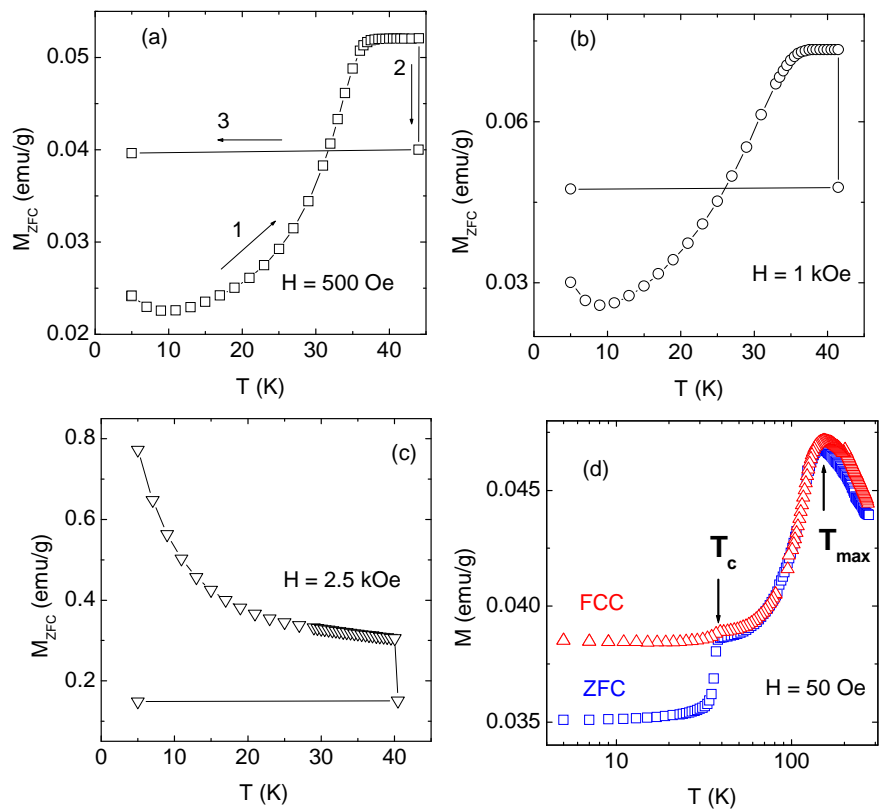